\documentclass[twocolumn,showpacs,preprintnumbers,amsmath,amssymb]{revtex4}

\usepackage{graphicx}
\usepackage{dcolumn}
\usepackage{bm}

\usepackage[usenames]{color}

\makeatletter
\def\lsim{\mathrel{\mathpalette\gls@align<}}
\def\gsim{\mathrel{\mathpalette\gls@align>}}
\def\gls@align#1#2{\lower.6ex\vbox
	{\baselineskip\z@skip\lineskip\z@\ialign
		{$\m@th#1\hfill##$\crcr#2\crcr\sim\crcr}}}
\makeatother

\def\be{\begin{equation}}
\def\ee{\end{equation}}
\def\bea{\begin{eqnarray}}
\def\eea{\end{eqnarray}}

\makeatletter
\def\leqq{\mathrel{\mathpalette\gle@align<}}
\def\geqq{\mathrel{\mathpalette\gle@align>}}
\def\gle@align#1#2{\lower.6ex\vbox{%
\baselineskip\z@skip\lineskip\z@\ialign{$\m@th#1\hfill##$\crcr#2\crcr=\crcr}}}
\makeatother

\begin{document}

\preprint{
Step Bunching with Alternation of  Structural Parameters
}

\title{
Step Bunching with Alternation of  Structural Parameters
}

\author{Masahide Sato}
	\altaffiliation[Present address:]{
	Imformation Media Center of Kanazawa University.}
	\email{sato@cs.s.kanazawa-u.ac.jp}
\author{Makio Uwaha$^a$}
\author{Tomonori Mori}
\author{Yukio Hirose}
\affiliation{
	Department of Computational Science, Kanazawa University,Kakuma-machi,
        Kanazawa 920-1192, Japan\\
	$^a$Department of Physics, Nagoya University,Furo-cho, 
        Chikusa-ku, Nagoya 464-8602, Japan \\
}

\begin{abstract}
By taking account of the alternation of structural parameters,
we study bunching of impermeable steps induced by drift of adatoms
on a vicinal face of Si(001).
With the alternation of diffusion coefficient,
the step bunching occurs irrespective of the direction of the drift 
if the step distance is large.
Like the bunching of permeable steps,
the type of large terraces is determined by the drift direction.
With step-down drift,
step bunches grows  faster than those with step-up drift.
The ratio of the growth rates is 
larger than 
the ratio of the diffusion coefficients.
Evaporation of adatoms, which does not cause the step bunching,
decreases the difference.
If only the alternation of kinetic coefficient is taken into account,
the step bunching occurs with step-down drift.
In an early stage,
the initial fluctuation of the step distance determines 
the type of large terraces,
but in a late stage,
the type of large terraces is
opposite to the case 
of alternating diffusion coefficient.
\end{abstract}


\maketitle

\section{Introduction}

When a  Si(001) vicinal face is tilted in the $ \langle 110 \rangle $ direction,
two types of terraces, $1 \times 2$ and $2 \times 1$ terraces,
appear alternately.
In the $1 \times 2 $ terrace, which we call  T$_\mathrm{A}$,
the surface diffusion perpendicular to the steps is faster than 
that parallel to the steps.
In the $2 \times 1$ terrace,  which we call  T$_\mathrm{B}$,
the relation is opposite.

These terraces are separated by single steps.
The type of the steps also changes alternately:
the step at the upper side edge of T$_\mathrm{B}$,
which we call $\mathrm{S}_\mathrm{B}$, is rougher than 
that of T$_\mathrm{A}$,
which we call $\mathrm{S}_\mathrm{A}$.
Then, parameters like kinetic coefficient and the step stiffness
may change for the two kinds of steps.

When a Si(001) vicinal face is heated by direct electric current,
the vicinal face is unstable and step bunching occurs
irrespective of the current direction~\cite{livin,latyshev-livin-aseev}.
The type of large terraces between bunches is
$\mathrm{T}_\mathrm{A}$ with step-down current
and $\mathrm{T}_\mathrm{B}$  with step-up current.
Cause of the step bunching is considered to be 
the drift of adatoms~\cite{Ichikawa-d92apl} induced by the current.
By using a one-dimensional step model,
where the alternation of diffusion coefficient
and that of kinetic coefficient are taken into account,
Stoyanov~\cite{stoyanov-si001} 
theoretically studied  the stability of a vicinal face  for pairing of steps.
With large kinetic coefficients,
the step pairing occurs irrespective of the direction of the drift,
and the type of large terraces between step pairs 
is determined by the current direction.
To study behaviors of step pairs,
Natori and co-workers carried out 
numerical simulation of a similar one-dimensional 
step model~\cite{natori-f.y,natori-f.f}.
With step-down drift,
the step bunching occurs via coalescence of step pairs,
but the step bunching does not occur with step-up drift,
which  disagrees with the experiments~\cite{livin,latyshev-livin-aseev}.

Recently,
we carried out Monte Carlo simulation~\cite{Sato-us02jcg},
where the alternation of anisotropic surface diffusion was taken into account.
In contrast to the previous studies~\cite{natori-f.y,natori-f.f},
the step bunching occurs irrespective of 
the direction of the drift 
as in the experiment~\cite{livin,latyshev-livin-aseev}.
In our model~\cite{Sato-us02jcg}, the steps are perfectly permeable,
and the alternation of kinetic coefficient and the evaporation
of adatoms are neglected.
On the other hand,
the steps are impermeable and
both the alternation of kinetic coefficient and the evaporation are taken
into account in the previous models~\cite{natori-f.y,natori-f.f}.
Since there are many differences between the models,
it is not clear what is the most important factor 
to cause the different results.

In this paper,
we use a one-dimensional model of impermeable steps
to study the drift-induced step bunching.
In Sec.~\ref{sec:model}, we introduce the model.
In Sec.~\ref{sec:diffusion-coefficient},
we analyze the model with alternating diffusion coefficient
and perform a numerical simulation.
We compare the results with our previous study~\cite{Sato-us02jcg}.
In Sec.~\ref{sec:step-kinetics},
we study the model with the alternating kinetic coefficient.
The effect of evaporation, which is neglected in Ref.~\onlinecite{Sato-us02jcg},
is also studied in Secs.~\ref{sec:diffusion-coefficient}
and \ref{sec:step-kinetics}.
In Sec.~\ref{sec:summary}
we summarize the results and give a brief discussion.

\section{Model}\label{sec:model}
We use a one-dimensional step flow model~\cite{stoyanov-si001,natori-f.y,natori-f.f}.
The $y$-coordinate is taken in the step-down direction.
When the drift of adatoms  is parallel to the $y$-axis,
the diffusion equation of adatom density $c(y,t)$ is given by
	\begin{equation}
	\frac{\partial c}{\partial t}
	=
	D_{m} \frac{\partial^2 c}{\partial y^2}
	-\frac{D_m F}{k_{\rm B}T} \frac{\partial c}{\partial y}
	-\frac{1}{\tau}c,
	\label{eq:diffusion}
	\end{equation}
where $D_{m}$ is the diffusion coefficient in the $m$th terrace,
$F$  the force to cause the drift
and $\tau$ the lifetime of adatoms for evaporation.

\begin{figure}[htp]
	\includegraphics[width=1.0\linewidth]{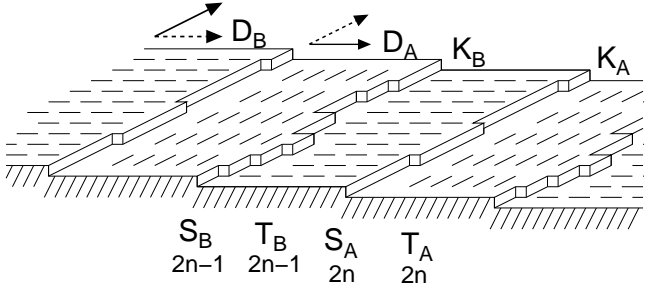}
	\caption{
	A restructed Si(001) vicinal face.
	Short lines represent dimers.
	}\label{fig:vicinal}
	\end{figure}

Boundary conditions at the $m$th step are 
given by~\cite{stoyanov-si001,Ozdemir-z92prb}
	\begin{eqnarray}
        K_{m}(\left. c \right|_{+}- c_{m})
	&=&
	D_{m}
	\left(
	\left. 	\frac{\partial c}{\partial y} \right|_{+}
	-\frac{F}{k_{\rm B}T}
	\left. 	c  \right|_{+}
	\right)
	\nonumber \\
	& &
	-P_m (\left. c \right|_+ - \left. c \right|_-),
	\label{eq:boundary1}
	\\
	K_{m}(\left. c \right|_{-}- c_{m})
	&=&
	-
	D_{m-1}
	\left(
	\left.
	\frac{\partial c}{\partial y}
	\right|_{-}
	-\frac{F}{k_{\rm B}T}
	\left.
	c
	\right|_{-}
	\right)
	\nonumber \\
	& &
	-P_m (\left. c \right|_- - \left. c \right|_+),
	\label{eq:boundary2}
	\end{eqnarray}
where $c_m$ is the equilibrium adatom density,
$K_m$ kinetic coefficient of the $m$th step
and  $+(-)$ indicates the lower (upper) side of the step.
$\mathrm{T}_\mathrm{A}$ and $\mathrm{T}_\mathrm{B}$
appear alternately,
and we assume  steps with odd numbers are $S_\mathrm{B}$ steps
and those  with even numbers are 
$S_\mathrm{A}$ steps (Figure~\ref{fig:vicinal}).
The diffusion coefficients and the kinetic coefficients are given by
$(D_{2n}, K_{2n}) =(D_\mathrm{A}, K_\mathrm{A})$
and $(D_{2n-1}, K_{2n-1}) = (D_\mathrm{B}, K_\mathrm{B})$.
Bearing a Si(001) vicinal face in mind,
we set $D_{A} >D_{B}$ and $K_{B} > K_{A}$.
The second terms in the right hand side of 
eqs.~(\ref{eq:boundary1}) and~(\ref{eq:boundary2})
represent the adatom current through the step without solidification.
With $P_m \to \infty$,
the difference between $\left. c \right|_{+}$ and $\left. c \right|_{-}$
vanishes and the step is called perfectly permeable.
With $P_m =0$, the step is called impermeable
and the surface diffusion field is separated by the step.
Hereafter we deal with the impermeable steps.

When the neighboring steps interact
with the potential $\zeta_m$,
the equilibrium adatom density at the $m$th step
$c_{m}$ is given by~\cite{mullins,Uwaha92}
	\begin{eqnarray}
	c_{m}
	&=& 
	c_{\rm eq}^0 
	 +\frac{\Omega c_{\rm eq}^0}{k_{\rm B}T}
	  \frac{\partial \zeta_m}{\partial y_{m}}
	\nonumber \\
	& &
	=
	c_\mathrm{eq}^0
	\left[
	1
	-
	\tilde{A}_{\tilde{\nu} }
	\left(
	 \frac{1}{l_{m-1}^{\tilde{\nu}} } -\frac{1}{l_{m}^{\tilde{\nu}}}
	 \right)
	\right]
	\label{eq:interaction}
	\end{eqnarray}
where $c_{\rm eq}^0$ is the equilibrium adatom density of an isolated step,
$\Omega $ the atomic area,
$y_{m}$ the position of the $m$th step
and 
$l_{m} =(y_{m+1} -y_{m})$ is the width of the $m$th terrace.
If the step interaction potential $\zeta_m $ is given by 
$\zeta_m = - A (\ln l_{m-1} + \ln l_{m})$ as in a Si(001) 
vicinal face~\cite{Alerhand-vmj88prl},
the exponent is  $\tilde{\nu}= 1$
and $\tilde{A}_{\tilde{\nu}}= \Omega A/k_{\rm B}T$.

By solving eq.~(\ref{eq:diffusion})
with the boundary conditions,
eqs.~(\ref{eq:boundary1}) and (\ref{eq:boundary2}),
in a quasi-static approximation ($\partial c /\partial t =0$),
we determine the adatom density.
The velocity of the $m$th step is given by
	\begin{eqnarray}
	V_{m}
	&=& 
	K_m (\left. c \right|_+ -c_m) + K_m (\left. c \right|_- -c_m)
	\nonumber \\
	& =& 
	\Omega
	\left(D_{m} 
	\left.
	\frac{\partial c}{\partial y}
	\right|_{+}
	-\frac{D_{m}F}{k_{\rm B}T}
	\left.
	c
	\right|_{+}
	\right)
	\nonumber \\
	& &
	-
	\Omega
	\left( D_{m-1}
	\left.
	\frac{\partial c}{\partial y}
	\right|_{-}
	-\frac{D_{m-1}F}{k_{\rm B}T}
	\left.
	c
	\right|_{-}
	\right).
	\end{eqnarray}
In the following,
to see how the step bunching changes with the alternation of parameters,
we separately study the effect of diffusion coefficients
and that with kinetic coefficients.

\section{Step bunching with alternation of the diffusion coefficient}
\label{sec:diffusion-coefficient}
We use the model of impermeable steps
to study the step bunching with alternation of diffusion coefficient.
We compare results wtih the previous study~\cite{Sato-us02jcg}
to see the effect of step permeability.
For simplicity, the alternation of kinetic coefficient is neglected.

\subsection{Step bunching induced by the drift 
in a conserved system}\label{sec:sbinc}
We first study the step bunching in a conserved system.
Without the evaporation,
the step velocity is given by
	\begin{eqnarray}
	V_{m}
	& =&
	\frac{
	\Omega K D_{m-1}f (c_{m-1} e^{f l_{m-1}} -c_m)
	}{
	(e^{fl_{m-1}}-1)K + (e^{fl_{m-1}}+1)D_{m-1}f
	}
	\nonumber \\
	&&
	-
	\frac{
	\Omega K D_{m}f (c_{m} e^{f l_{m}} -c_{m+1})
	}{
	(e^{fl_{m}}-1)K + (e^{fl_{m}}+1)D_{m}f
	},
	\label{eq:step-velocity2}
	\end{eqnarray}
where $f =F/k_{\rm B}T$. 

In a vicinal face with the step distance $l$,
the step velocities are given by
\begin{eqnarray}
V_{2n}&=& -V_{2n-1}
	\nonumber \\
	& =&
	\frac{\Omega K^2 f c_\mathrm{eq}^0 (e^{fl}-1)^2
	(D_\mathrm{B}-D_\mathrm{A})
	}{
	v_A v_B
	},
	\label{eq:v2m}
\end{eqnarray}
where $v_A$ and $v_B$ are
	\begin{eqnarray}
	v_A &=&
	(e^{fl} -1)K + (e^{fl}+1)D_\mathrm{A} f,
	\nonumber \\
	v_B &=&
	(e^{fl} -1)K + (e^{fl}+1)D_\mathrm{B} f.
	\end{eqnarray}
Since the diffusion coefficient $D_\mathrm{A}$ is larger than $D_\mathrm{B}$,
$S_\mathrm{A}$ steps advance and $S_\mathrm{B}$ steps recede
with step-down drift ($f >0$).
With step-up drift ($f <0$), 
the direction of the step motion is reversed.

In the initial stage of the instability,
formation of large $\mathrm{T_A} (\mathrm{T_B})$ terrace with 
step-down (step-up) drift is expected due to the step pairing.
To study the motion of pairs of steps,
we carry out numerical integration of eq.~(\ref{eq:step-velocity2}).
In addition to the step distance $l$,
there are two characteristic lengths
in the vicinal face.
One is $f^{-1} \sim k_{\rm B}T/e E$~\cite{yagi}, 
which  is a characteristic length determined by
the drift and usually much larger than the step distance.
The other  is $D_\mathrm{A}/K$,
which is characteristic length determined by step kinetics.
When the step distance $l$ is much smaller than $D_\mathrm{A}/K$,
the step kinetics is more important than the surface diffusion.
We assume that $f^{-1}$ is larger than $D_\mathrm{A}/K$ and $l$,
and study the time evolution in two cases;
$D_\mathrm{A}/K <l$ and $l <D_\mathrm{A}/K$.

	\begin{figure}[htp]
	\includegraphics[width=0.7\linewidth]{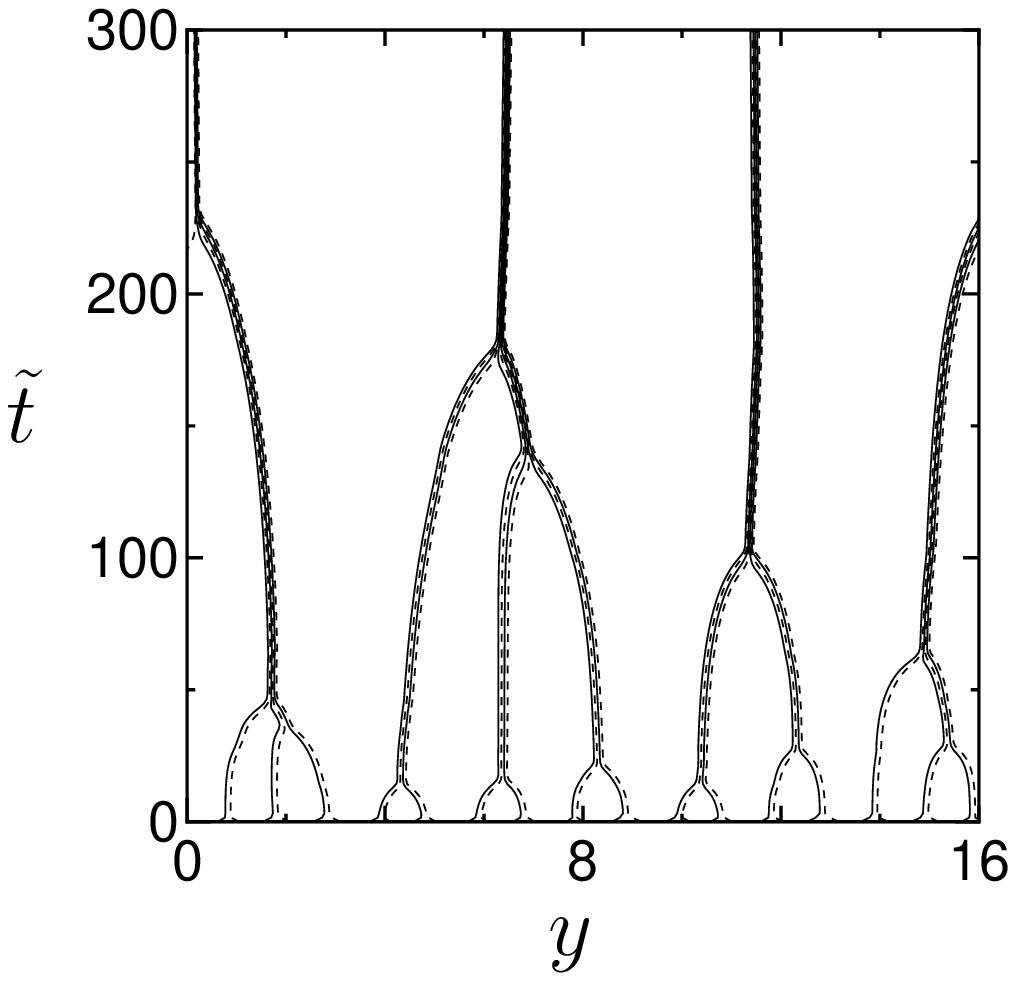}
	\centerline{\hspace*{1.0cm} {\sf \large (a)}}
	\vspace*{0.2cm}
	\\
	\includegraphics[width=0.7\linewidth]{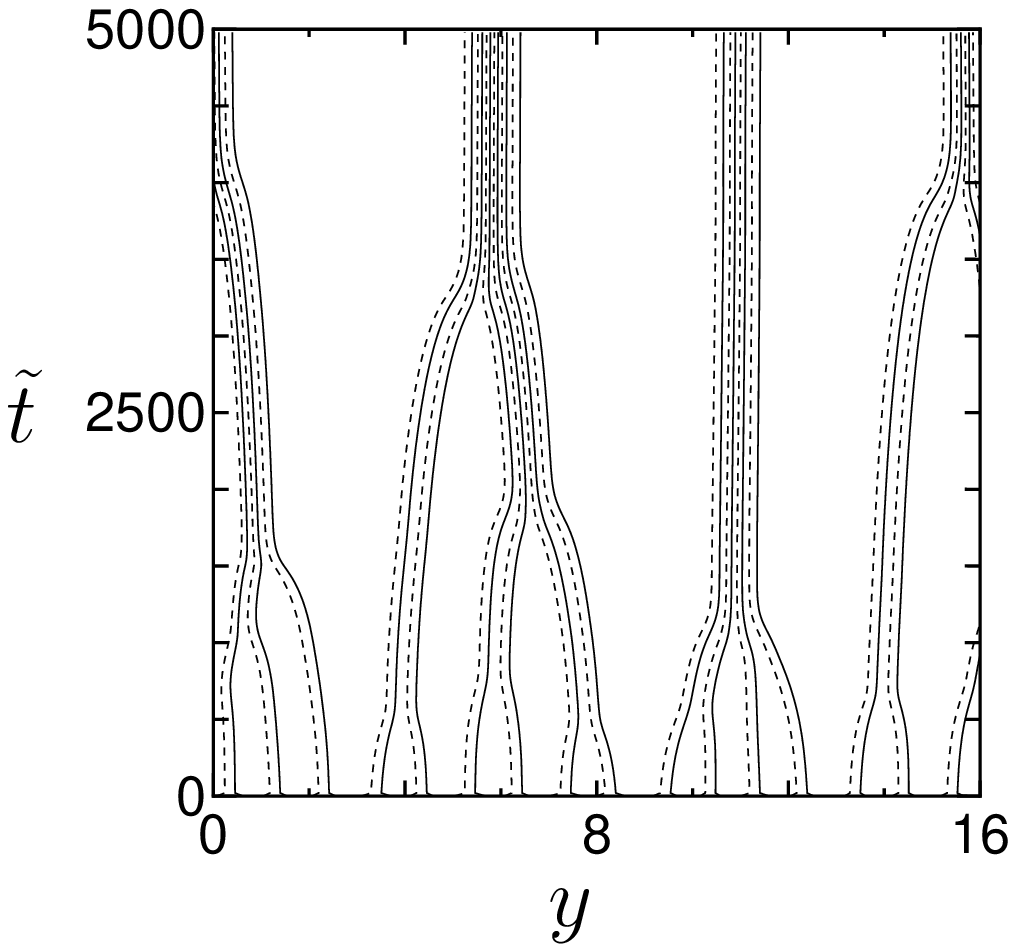}
	\centerline{\hspace*{1.0cm} {\sf \large (b)}}
	\caption{
	Time evolution of step position.
	The step distance $l$ is larger than $D_\mathrm{A}/K$
	and $ D_\mathrm{B}/K$.
	The drift direction is (a) step-down drift
	and (b) step-up,
	and the drift velocities satisfy $|fl|=  0.1$.
	The number of steps is 32 and the system size is 16 with
	 the periodic boundary condition. 
	$\tilde{A}_{\tilde{\nu}} =5 \times 10^{-3}$ with $\tilde{\nu} =1$,
	$D_{\mathrm{A}}/Kl =0.2$, $D_{B}/Kl =0.02$
	and $\tilde{t} \equiv \Omega c_\mathrm{eq}^0 t$.
	}\label{fig:evolution}
	\end{figure}

Figure~\ref{fig:evolution} represents
the time evolution of step positions with
 $D_{\mathrm{A}}/K \ll l \ll f^{-1}$.
This is  the case in which the diffusion is slow and controls
the time evolution.
Solid lines represent the evolution of the positions of $\mathrm{S}_\mathrm{B}$
and dotted lines represent  S$_{\rm A}$.
The characteristic length $f^{-1}$
is  $ f^{-1} = 200$ and 
the scaled time $\tilde{t}$ is given by $\tilde{t} \equiv \Omega c_\mathrm{eq}^0 t$.
The initial step distance is about the same,
but  with a random fluctuation.
In the initial stage, the step pairing occurs irrespective of the drift direction.
Large terraces are $\mathrm{T}_\mathrm{A}$  with step-down 
drift (Fig.~\ref{fig:evolution}(a))
and $\mathrm{T}_\mathrm{B}$ with step-up drift (Fig.~\ref{fig:evolution}(b)),
which agrees with eq.~(\ref{eq:v2m}) 
and the previous analysis~\cite{stoyanov-si001,natori-f.y,natori-f.f}.

In a late stage,
large bunches appear irrespective of the drift direction.
The type of large terraces is the same as in the initial stage.
Free single steps or pairs do not exist on the large terraces
and the bunches grow via coalescence of small bunches.
Though the formation process of large bunches is
similar in both cases,
the step density with step-down drift
is higher than that with step-up drift,
and  the growth is much faster.
The ratio of the growth rate of step bunches is 
larger than the ratio of the diffusion coefficients.
In the numerical study of Natori et al.~\cite{natori-f.y,natori-f.f},
the formation of large bunches with step-up drift did not occur
in contrast to our simulation.
For the very slow growth rate of step bunches,
their simulation time might be too short to produce large bunches.

	\begin{figure}[htp]
	\includegraphics[width=0.7\linewidth]{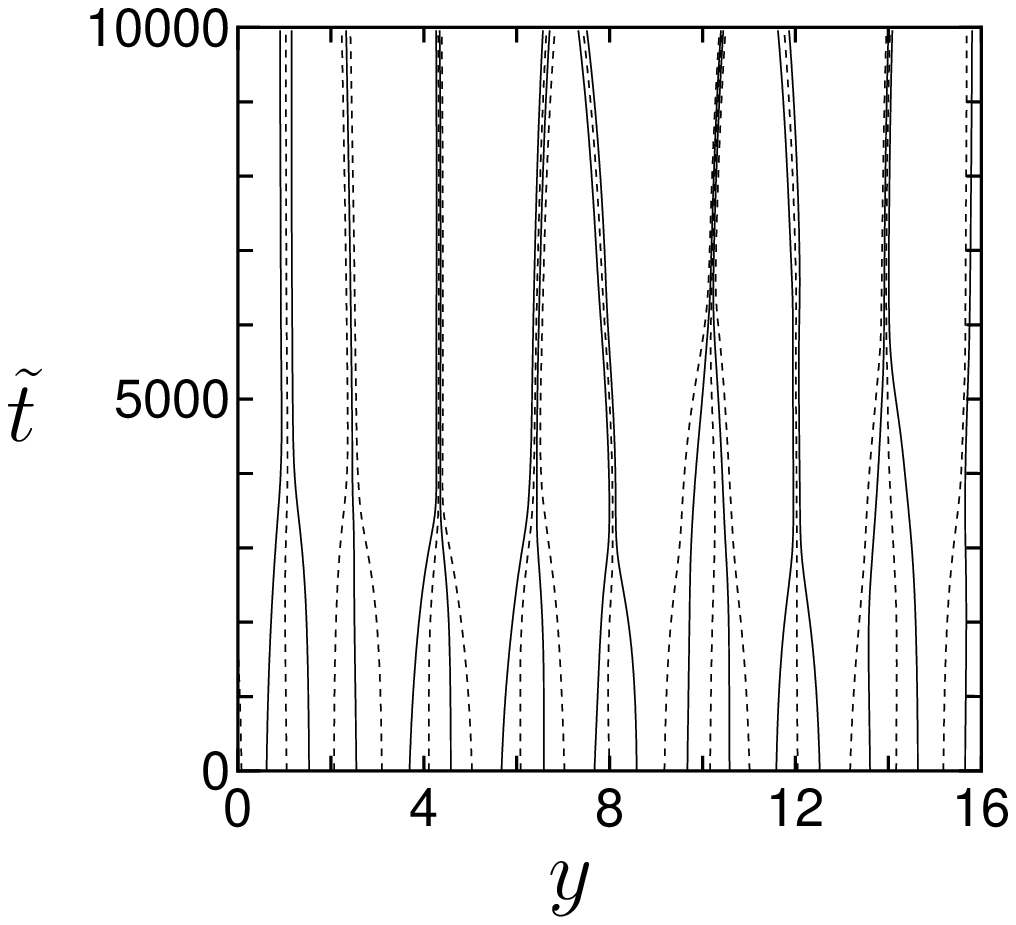}
	\centerline{\hspace*{1.0cm} {\sf \large (a)}}
	\vspace*{0.2cm}
	\\
	\includegraphics[width=0.72\linewidth]{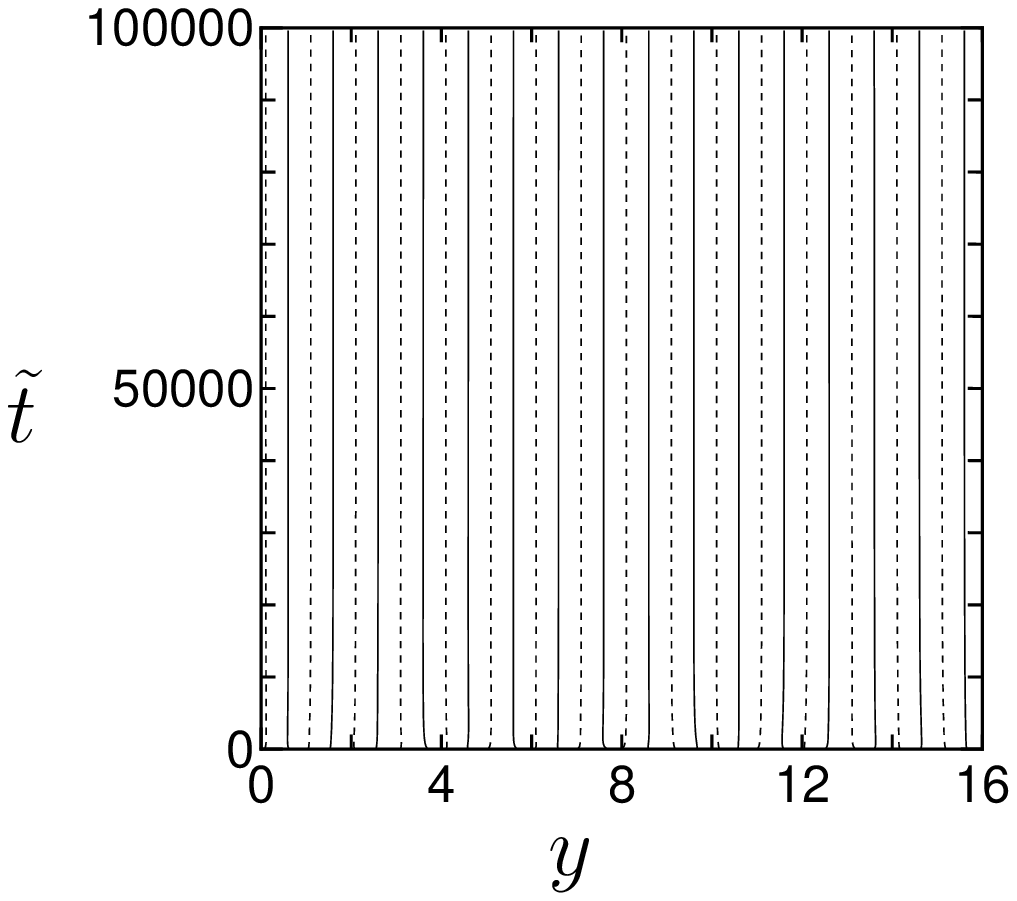}
	\centerline{\hspace*{1.0cm} {\sf \large (b)}}
	\caption{
	Time evolution of step position 
  	with step distance $l \ll D_{\mathrm{A}}/K, D_\mathrm{B}/K$
	with 
	(a) step-down drift and
	(b) step-up drift.
	The number of steps is 32
	and the system size is 16 with the periodic
	boundary condition. 
	$\tilde{A}_{\tilde{\nu}} =4 \times 10^{-3}$ with $\tilde{\nu} =1$,
	$|fl|$=0.1, $D_\mathrm{A}/Kl =200$
	and $D_{\mathrm{B}}/Kl =20$.
	}\label{fig:short-alternation-diffusion}
	\end{figure}

Figure~\ref{fig:short-alternation-diffusion} represents
the step bunching with $l \ll D_B/K$,
{\it i.e.}, the step kinetics-contral case.
The amplitude of the initial fluctuation is the same as
that in Fig.~\ref{fig:evolution}.
When the drift is in the step-down direction
(Fig.~\ref{fig:short-alternation-diffusion}(a)),
the equidistant step train is unstable and the step bunching occurs.
The terrace type between step bunches is determined
by the initial fluctuation and both types of large terraces coexist.
When the drift is in the step-up direction
(Fig.~\ref{fig:short-alternation-diffusion}(b)),
the step bunching does not seem to occur.
The results  are very different from Fig.~\ref{fig:evolution}.

	\begin{figure}[htp]
	\includegraphics[width=0.7\linewidth]{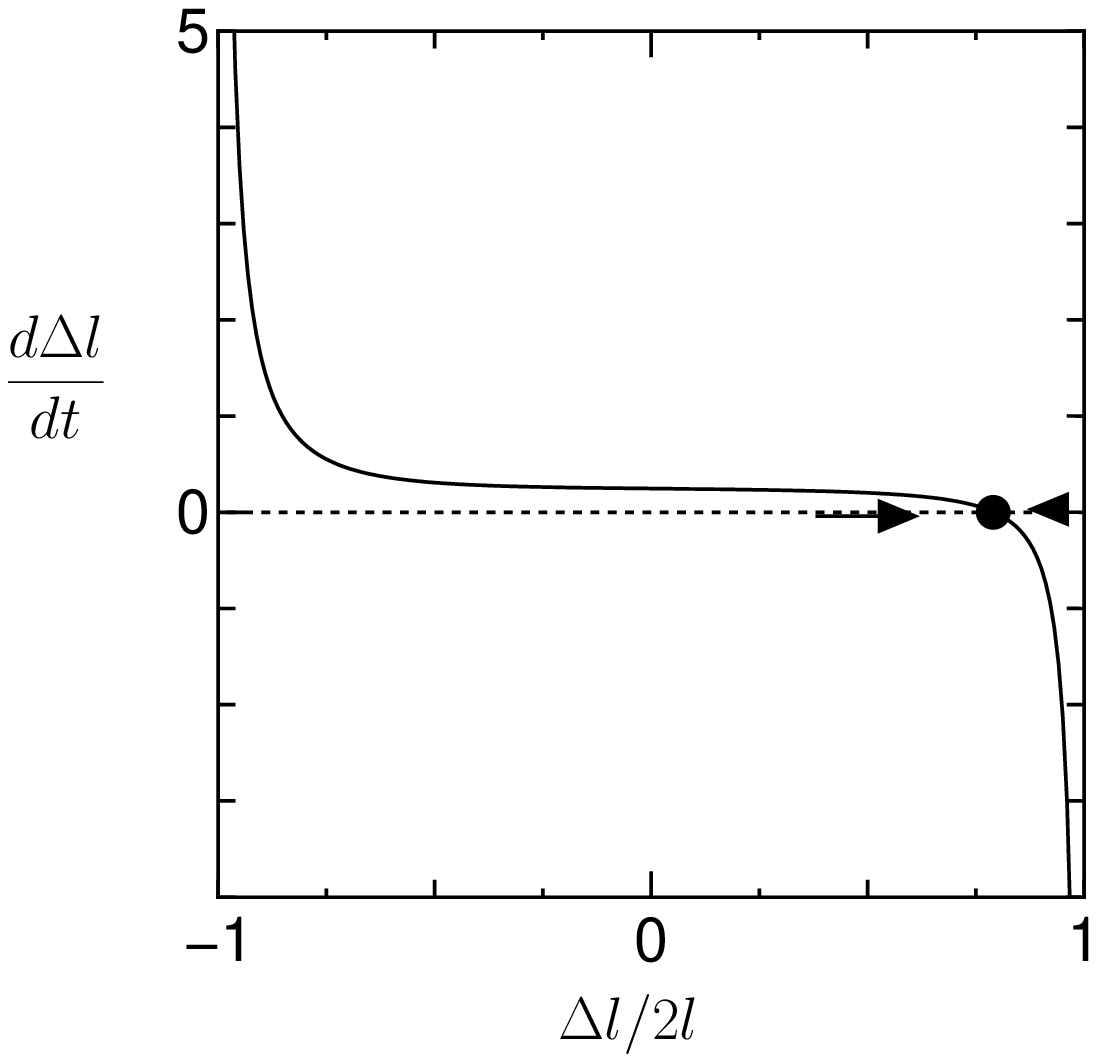}
	\centerline{\hspace*{1.0cm} {\sf \large (a)}}
	\vspace*{0.2cm}
	\\
	\includegraphics[width=0.72\linewidth]{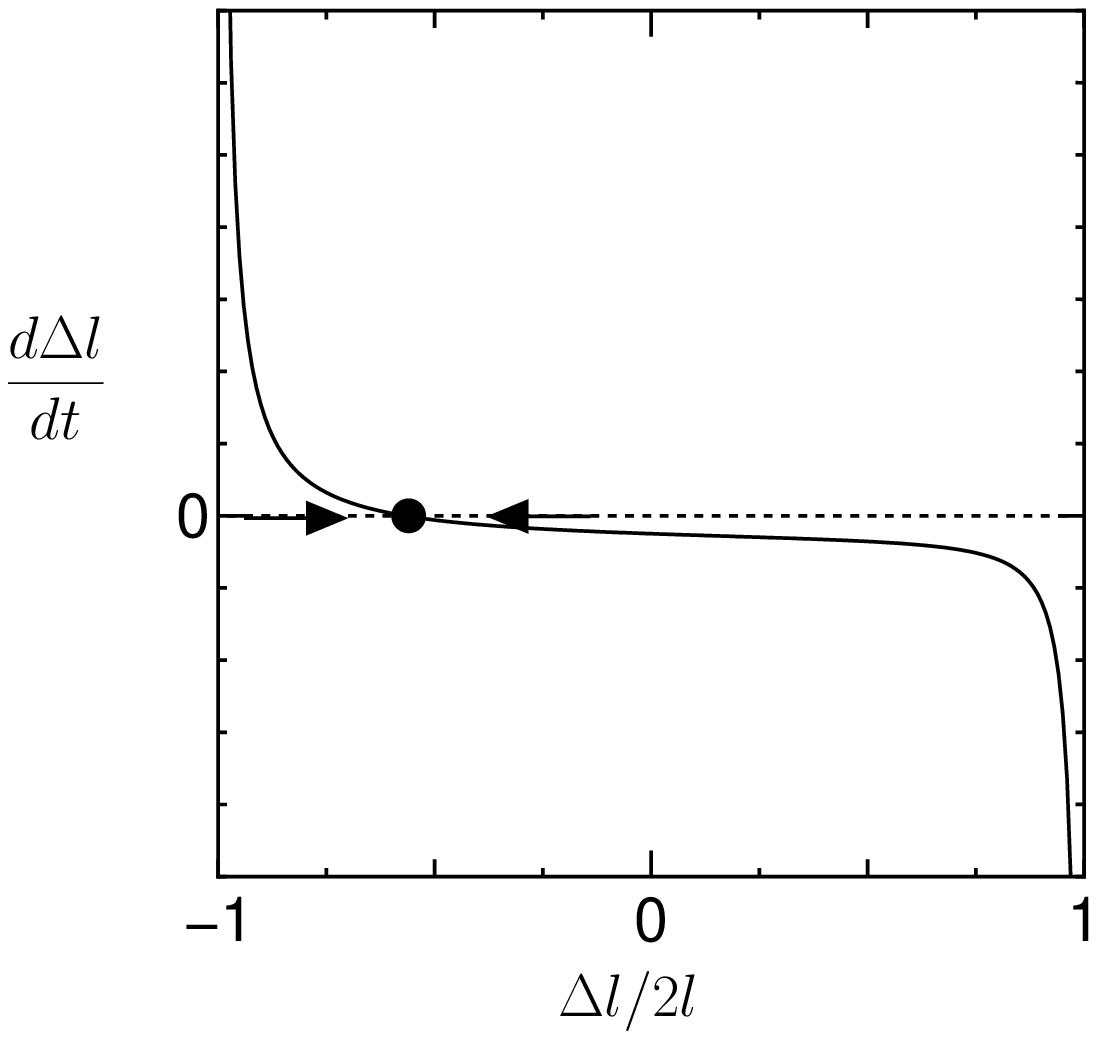}
	\centerline{\hspace*{1.0cm} {\sf \large (b)}}
	\caption{
	Time derivative of the terrace width change $\Delta l$
	with (a) step-down drift and (b) step-up drift.
	Parameters are the same as in Fig.~\ref{fig:evolution}.
	}\label{fig:short2}
	\end{figure}

To find the reason that the step behavior changes with the kinetic coefficients,
we analyse the change of alternating terrace width.
We assume the width of  $T_\mathrm{A}$ is 
$l_{2n}=l+\Delta l/2$ and that of $T_\mathrm{B}$ is $l_{2n-1}=l-\Delta l/2$.
Figure~\ref{fig:short2} represents
the time derivative of $\Delta l$ for 
large kinetic coefficients.
Parameters are the same as in Fig.~\ref{fig:evolution}.
With step-down drift (Fig.~\ref{fig:short2}(a)),
the initial vicinal face $\Delta l =0$ is unstable
and $\Delta l$ increases up to the stable fixed point near $\Delta l=2l$. 
The surface consists of large $T_\mathrm{A}$ terraces and small
$T_\mathrm{B}$ terraces, {\it i.e.}, tight step pairs.
With step-up  drift (Fig.~\ref{fig:short2}(b)),
the initial stage is also unstable and $\Delta l$ decreases to the stable fixed
point near $\Delta l= -2l$.
The surface consists of small $T_\mathrm{A}$ terraces and 
large $T_\mathrm{B}$ terraces.
The results agree with the initial pairing stage of 
Fig.~\ref{fig:evolution}.

	\begin{figure}[htp]
	\includegraphics[width=0.7\linewidth]{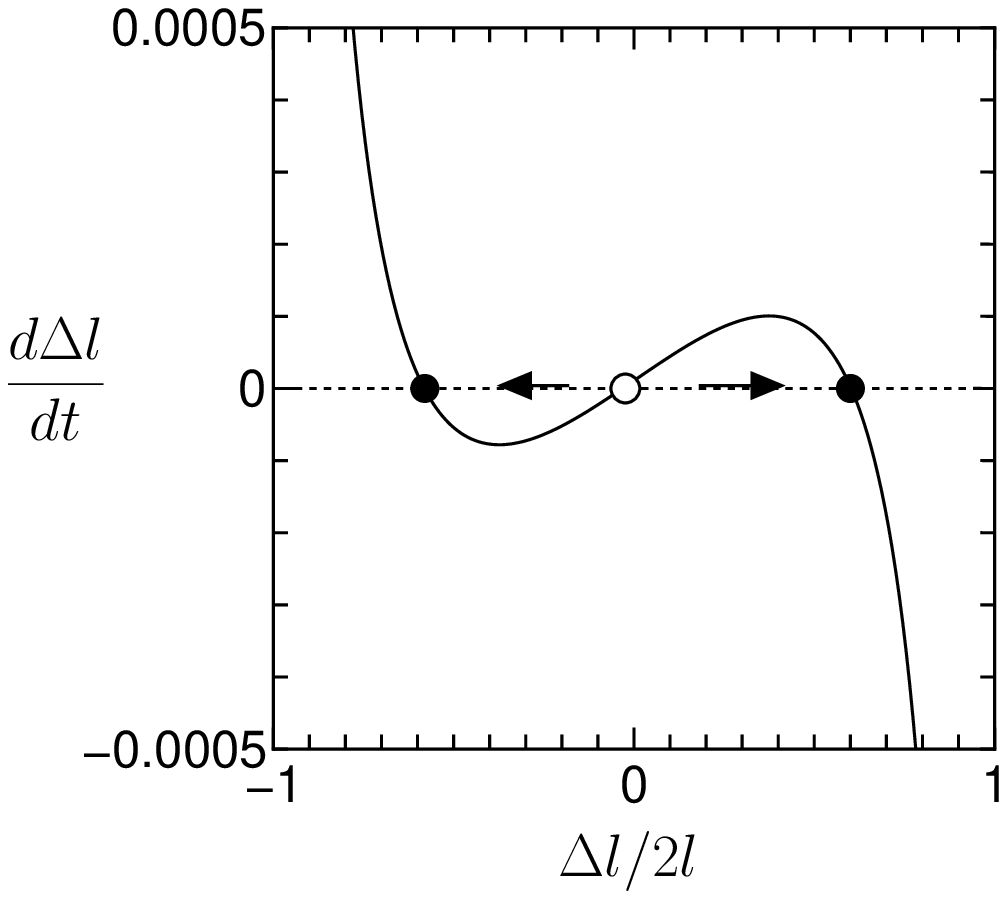}
	\centerline{\hspace*{1.0cm} {\sf \large (a)}}
	\vspace*{0.2cm}
	\\
	\includegraphics[width=0.72\linewidth]{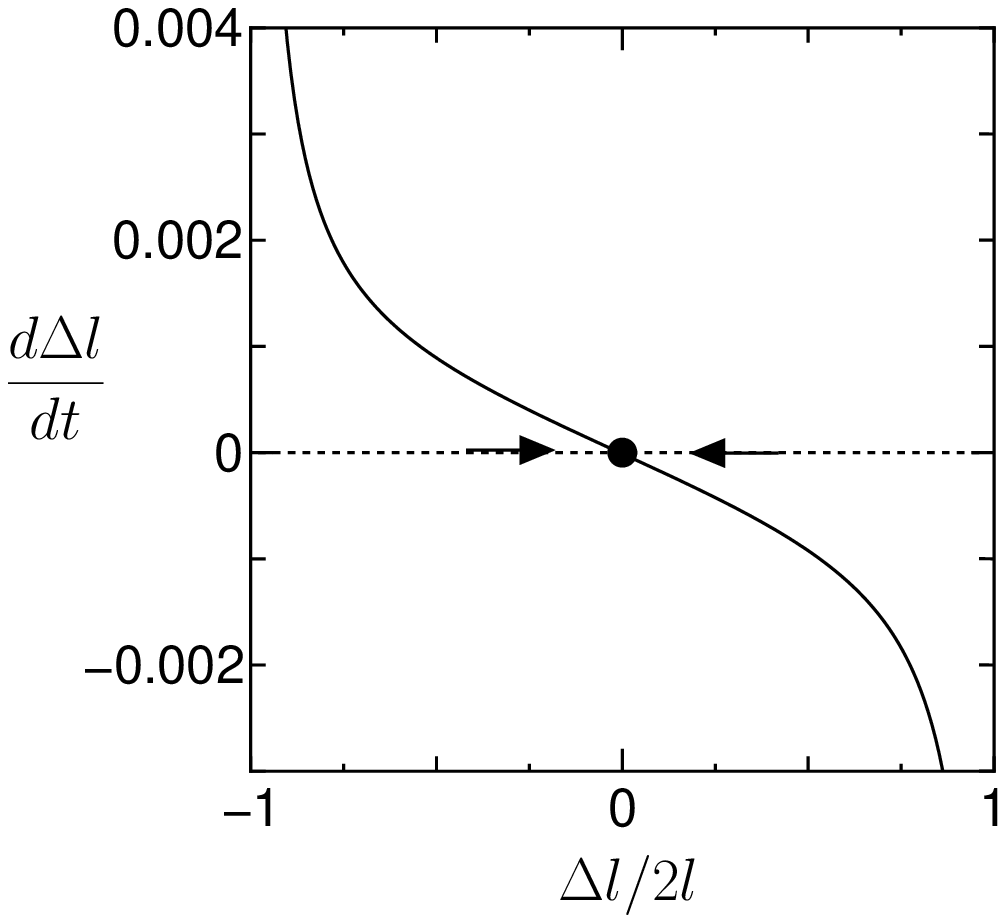}
	\centerline{\hspace*{1.0cm} {\sf \large (b)}}
	\caption{
	Time derivative of the terrace width change $\Delta l$
	with (a) step-down drift and (b) step-up drift.
	Parameters are the same as in Fig.~\ref{fig:short-alternation-diffusion}.
	}\label{fig:short1}
	\end{figure}

Figure~\ref{fig:short1}
represents the time derivative of $\Delta l$ for 
small kinetic coefficients.
With step-down drift (Fig.~\ref{fig:short1}(a)),
there are three fixed points.
The fixed point (open circle) near the center is unstable 
and two other fixed points (filled circles) are stable.
Since the unstable fixed point is very close to $\Delta l=0$,
the final stage is not unique if the initial fluctuation is included.
In Fig.~\ref{fig:short-alternation-diffusion}(a),
the initial random fluctuation is not negligible and 
some terraces move to the fixed point with positive $\Delta l$
and two types of terraces coexist.

With step-up drift,
there is only one stable fixed point  with a small negative  $\Delta l $.
From the condition $V_m$ in eq.~(\ref{eq:step-velocity2}),
the difference of step distance $\Delta l$ 
in the fixed point near $\Delta l = 0$ is given by
	\begin{equation}
	\frac{\Delta l}{l}
	= \frac{Kl(D_B-D_A)}{2D_{A}D_{B}},
	\label{eq:dlc1}
	\end{equation}
where we neglected the step-step repulsive interaction.
The sign of $\Delta l$ is determined by the diffusion coefficients
and independent of the drift direction as seen in
Fig.~\ref{fig:short-alternation-diffusion}(b).

\subsection{Step bunching induced by the evaporation}

Since the experiments~\cite{livin,latyshev-livin-aseev}
were  carried out at high temperatures,
the evaporation of adatoms may not be negligible for the step bunching.

We first neglect the drift of adatoms
and see if the step bunching occurs with the evaporation.
For simplicity,
we consider the limit of the fast step kinetics, $K \to \infty$.
If the step repulsion is absent,
the step velocity is given by
	\begin{equation}
	 \frac{V_{m}}{\Omega c_\mathrm{eq}^0}
	=
	-
	  \frac{D_{m}}{x_{m}}
		\tanh \frac{l_{m}}{2 x_{m}}
	-
	  \frac{D_{m-1}}{x_{m-1}}
		\tanh \frac{l_{m-1}}{2 x_{m-1}},
	\label{eq:velocity-vel3}
	\end{equation}
where $x_{m} = \sqrt{D_{m} \tau}$ is the surface diffusion length
in the $m$th terrace.
With the same step distance $l$,
the step velocity  is given by
	\begin{eqnarray}
	V_{2n}
	 &=& V_{2n-1}
\nonumber \\
	&=&
	-\Omega c_\mathrm{eq}^0
	\left[
	\frac{D_\mathrm{A} }{x_\mathrm{A}} \tanh \frac{l}{2x_\mathrm{A}} 
	+\frac{D_\mathrm{B}}{x_\mathrm{B}} \tanh \frac{l}{2x_\mathrm{B}}
	\right],
	\end{eqnarray}
where $x_\mathrm{A}$($x_\mathrm{B}$) represents
the surface diffusion length in $T_\mathrm{A}$($T_\mathrm{B}$)
and the  equidistant train of steps is a steady state.
When the terrace widths change alternately
and are given  by $l_{2n} = l + \Delta l/2$ and $l_{2n-1}=l- \Delta l/2$,
from eq.~(\ref{eq:velocity-vel3})
the time evolution of $\Delta l$
is given by
	\begin{equation}
	 \frac{1}{2}
	\frac{d \Delta l}{d t}
	 = V_{2n}- V_{2n-1} =0.
	\end{equation}
The vicinal face is marginal to the perturbation.
If the repulsive interaction is taken into account,
the vicinal face is stable and the step pairing does not occur.
With the alternation of diffusion coefficient,
the evaporation alone does not cause the step bunching.

When both the drift and the evaporation are present,
the step velocity is given by
	\begin{eqnarray}
	\frac{V_m}{\Omega}
	&=&
	\frac{(D_{m-1}-D_{m})fc_{m}}{2}
	\nonumber
	\\
	&-&
	\frac{
	D_{m-1} \alpha_{m-1} (c_m \cosh \alpha_{m-1} 
	l_{m-1} -e^{-fl_{m-1}/2} c_{m-1})
	 }{\sinh \alpha_{m-1} l_{m-1}}
	\nonumber 
	\\
	&-&
	\frac{
	D_m \alpha_m (c_m \cosh \alpha_m l_m -e^{-fl_m/2} c_{m+1}),
	 }{\sinh \alpha_m l_m}.
	\label{eq:velocity}
	\end{eqnarray}
The parameter $\alpha_m$ is defined by
	\begin{equation}
	\alpha_m = \frac{1}{2}\sqrt{f^2 + \frac{4}{x_m^2}}.
	\end{equation}
$\alpha_m^{-1}$ is the characteristic length of the diffusion
field in the $m$th terrace.

	\begin{figure}[htp]
	\includegraphics[width=0.7\linewidth]{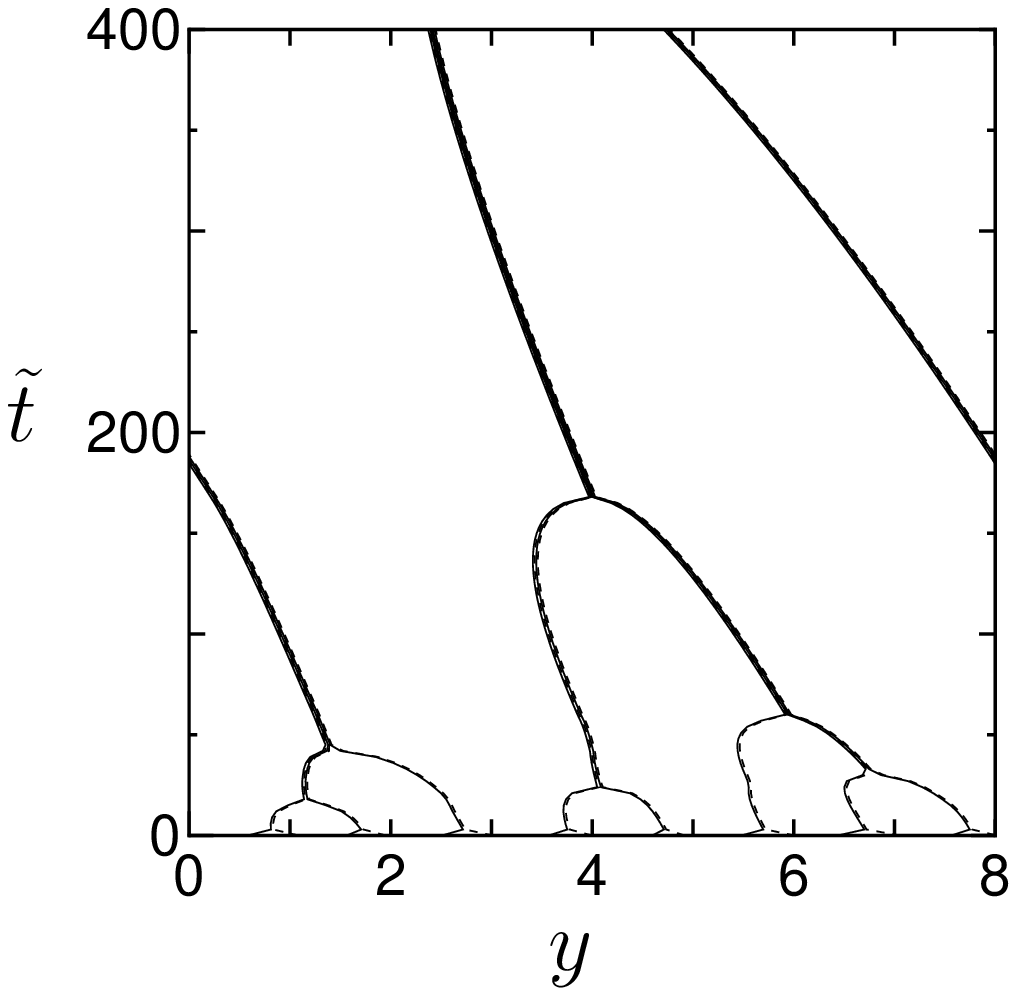}
	\centerline{\hspace*{1.0cm} {\sf \large (a)}}
	\vspace*{0.2cm}
	\includegraphics[width=0.7\linewidth]{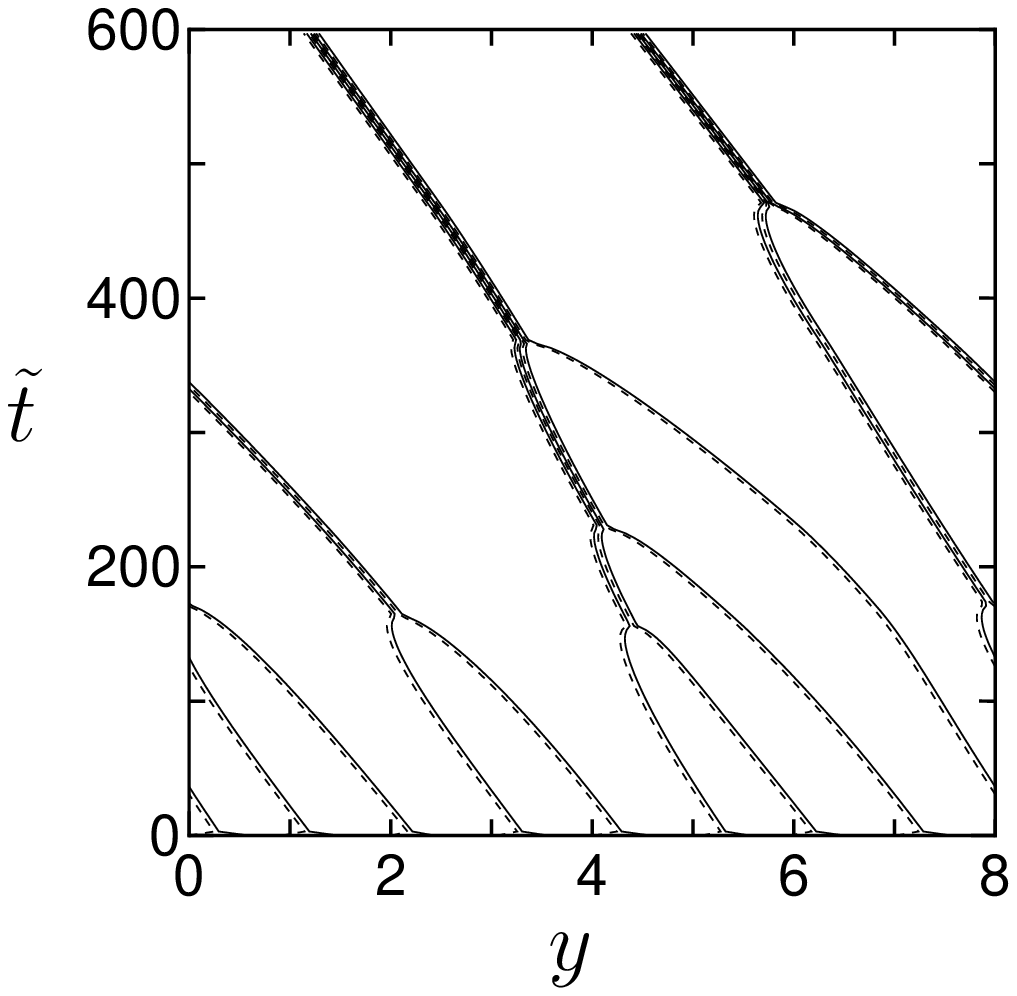}
	\centerline{\hspace*{1.0cm} {\sf \large (b)}}
	\caption{
	Time evolution of step position with evaporation
	with (a) step-down drift and (b) step-up drift.
	The number of steps is 16 and system size is 8.
	The diffusion coefficients are 
	$D_{B} =2$ and $D_{A}=10$.
	Other parameters are 
	$f =0.1$ 
	$\nu \Omega A/k_\mathrm{B}T = 10^{-4}$ with $\nu=1$,
	and $\tau =50$.
	}\label{fig:evolv-eva}
	\end{figure}

We carry out numerical integration of eq.~(\ref{eq:velocity}).
The time evolution of step positions (Figure~\ref{fig:evolv-eva})
shows the drift-induced step bunching with the evaporation,
where the characteristic length scale
in $\mathrm{T_A} $ is $\alpha_{2n}^{-1}= 14.9$ and 
that in $\mathrm{T_B}$ is $\alpha_{2n-1}^{-1}= 8.9$.
Irrespective of the drift direction,
very fast pairing of receding steps occurs at  the initial stage.
The type of large terraces is the same as in Fig~\ref{fig:evolution}
and not changed by the evaporation.

Large bunches appear by coalescence of small bunches.
With step-down drift,
coalescence of step bunches  occurs successively,
but the time interval of the coalescence increases 
with increasing the terrace width
between bunches (Fig.~\ref{fig:evolv-eva}(a)).
With step-up drift,
coalescence of step pairs does not occur until $\tilde{t} \approx 150$.
However, once the coalescence of step pairs starts,
the interval does not seem to increase much (Fig.~\ref{fig:evolv-eva}(b)).
The difference of the growth rate of bunches is much smaller
than that without the evaporation.

\section{Step bunching with alternation of kinetic coefficient}
\label{sec:step-kinetics}
In a Si(001) vicinal face,
not only diffusion coefficients but also the type of steps changes alternately.
In this section, 
we study the step bunching due to  the alternation of kinetic coefficient
and compare the result with Sec.~\ref{sec:diffusion-coefficient}.

\subsection{Step bunching induced by the drift}
To focus on the alternation of step kinetics,
we neglect the alternation of diffusion coefficients.
First, we neglect the evaporation.
Without the evaporation, the step velocity is given by
	\begin{eqnarray}
	V_{m}
	&=&
	\frac{
	\Omega D_{\rm s}f
	K_{m}K_{m-1} (e^{fl_{m-1}} c_{m-1} -c_{m})
	}{
	K_{m-1}(D_{\rm s} f -K_{m}) + K_{m} (D_{\rm s} f + K_{m-1})e^{fl_{m-1}}
	}
\nonumber \\
	&-&
	\frac{
	\Omega D_{\rm s}f
	K_{m}K_{m+1} (e^{fl_{m}} c_{m} -c_{m+1})
	}{
	K_{m}(D_{\rm s} f -K_{m+1}) + K_{m+1} (D_{\rm s} f + K_{m})e^{fl_{m}}
	},
	\label{eq:vmful}
	\end{eqnarray}
where  $D_\mathrm{s}$ is the diffusion coefficient.

	\begin{figure}
	\includegraphics[width=0.7\linewidth]{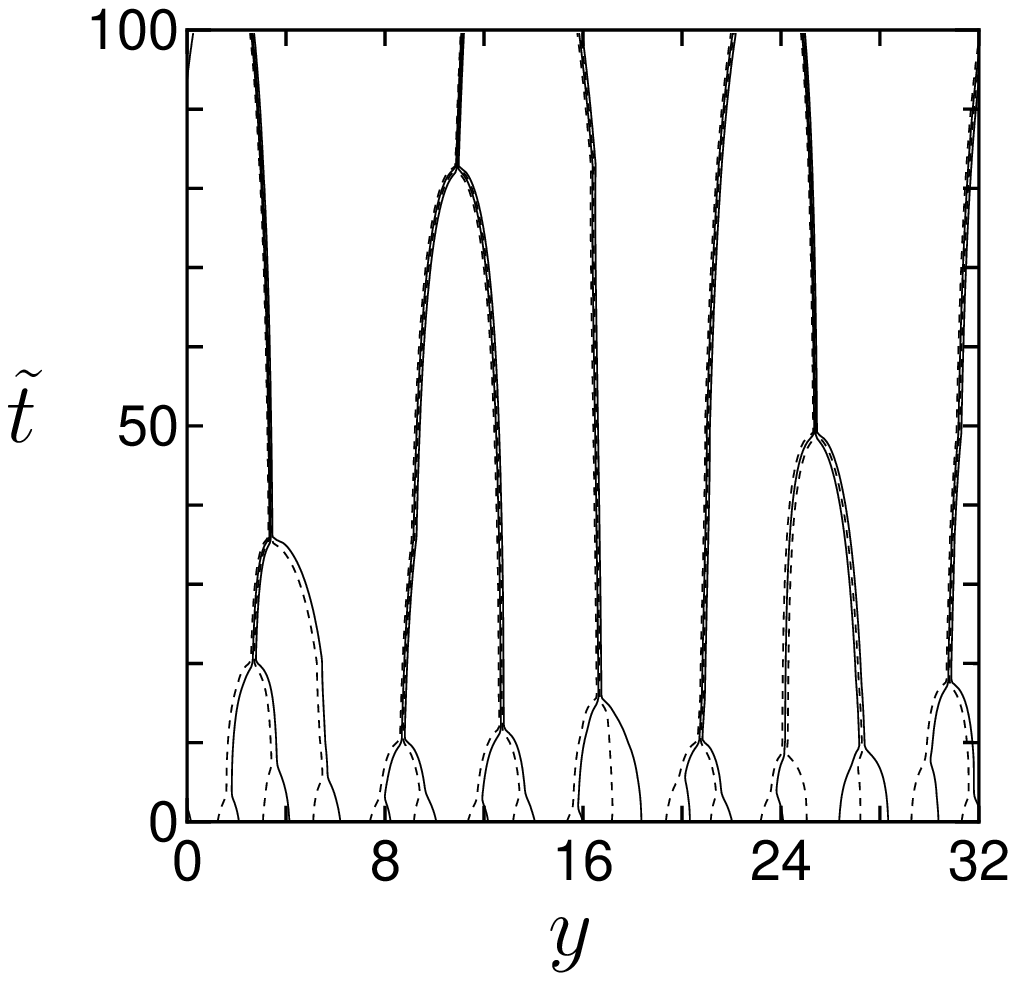}
	\centerline{\hspace*{1.2cm} {\sf \large (a)}}
	\vspace*{0.2cm}
	\\
	\includegraphics[width=0.72\linewidth]{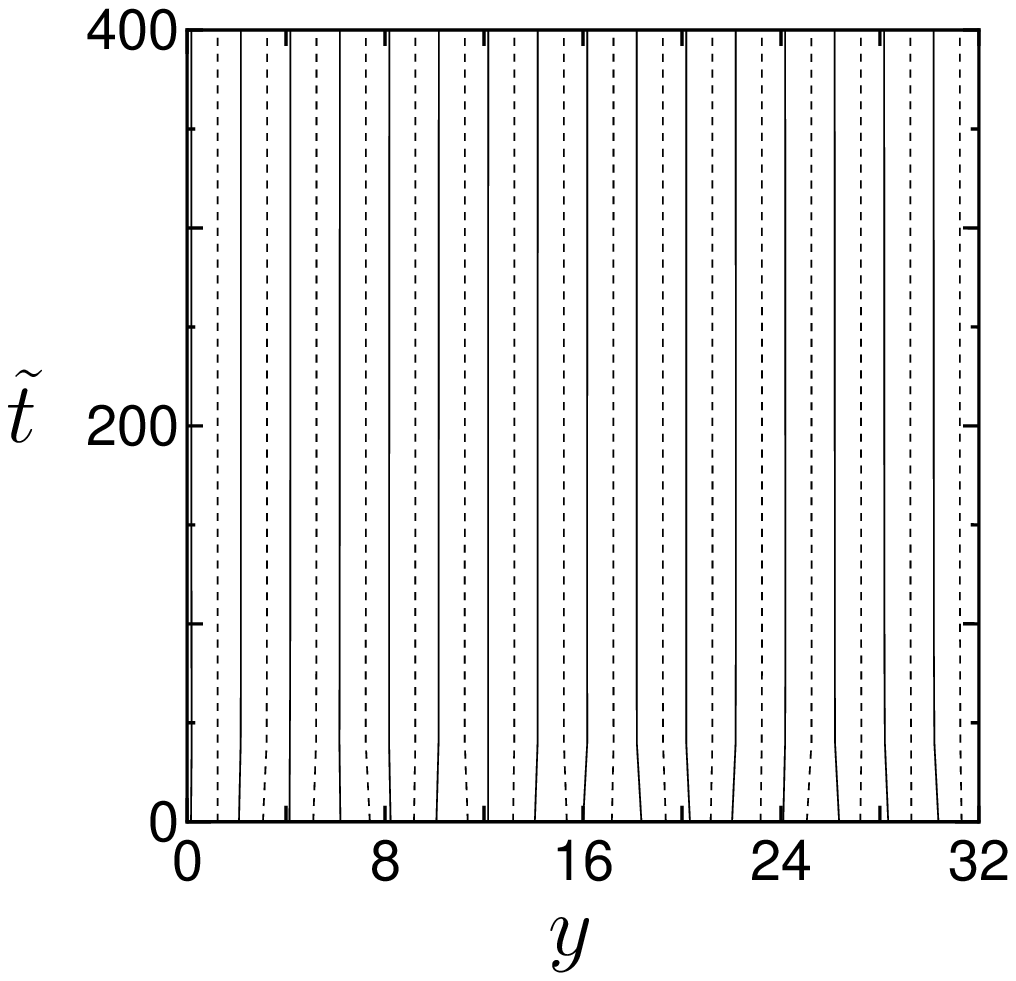}
	\centerline{\hspace*{1.2cm} {\sf \large (b)}}
	\caption{
	Time evolution of step position with alternating change 
	of kinetic coefficient with
	step distance $D_{\rm s}/K_\mathrm{A} \ll 1$.
	Drift is in 
	(a) step-down direction and
	(b) the step-up direction.
	The number of steps is 32
	and
	the system size is 32 with the periodic boundary condition.
	Parameters are 
	$\tilde{A}_{\tilde{\nu}} =2 \times 10^{-2}$ with $\tilde{\nu}=1$,
	$|fl|=0.2$, $D_s/K_\mathrm{A} l =  1.0 \times 10^{-2}$
	and $D_{s}/K_\mathrm{B} l = 1.0 \times 10^{-3}$.
	}\label{fig:alternating-kinetics1}
	\end{figure}

In a vicinal face with equidistant steps,
the step velocities are given by
	\begin{eqnarray}
	V_{2n} 
	&=& -V_{2n-1}
	\nonumber \\
	&=&
	\frac{\Omega D_s K_A K_B c_\mathrm{eq}^0 f^2 
	(e^{fl}-1)^2 (K_B -K_A)
	}{
	g_\mathrm{A}g_\mathrm{B}
	},
	\label{eq:vvv}
	\end{eqnarray}
where $g_\mathrm{A}$ and $g_\mathrm{B}$ are 
	\begin{eqnarray}
	g_\mathrm{A}
	&=&
	[K_{A}(D_{\rm s} f -K_{B}) + K_{B} (D_{\rm s} f + K_{A})e^{fl}],
	\nonumber \\
	g_\mathrm{B}
	&=&
	[K_{B}(D_{\rm s} f -K_{A}) + K_{A} (D_{\rm s} f + K_{B})e^{fl}].
	\end{eqnarray}
Since the kinetic coefficient $K_B $ is larger than $K_A$,
$S_\mathrm{A}$ steps advance and $S_\mathrm{B}$ steps recede.
Alternation of large $\mathrm{T_A}$ terraces
and small $\mathrm{T_B}$ terraces is expected by formation of step pairs.

	\begin{figure}
	\includegraphics[width=0.7\linewidth]{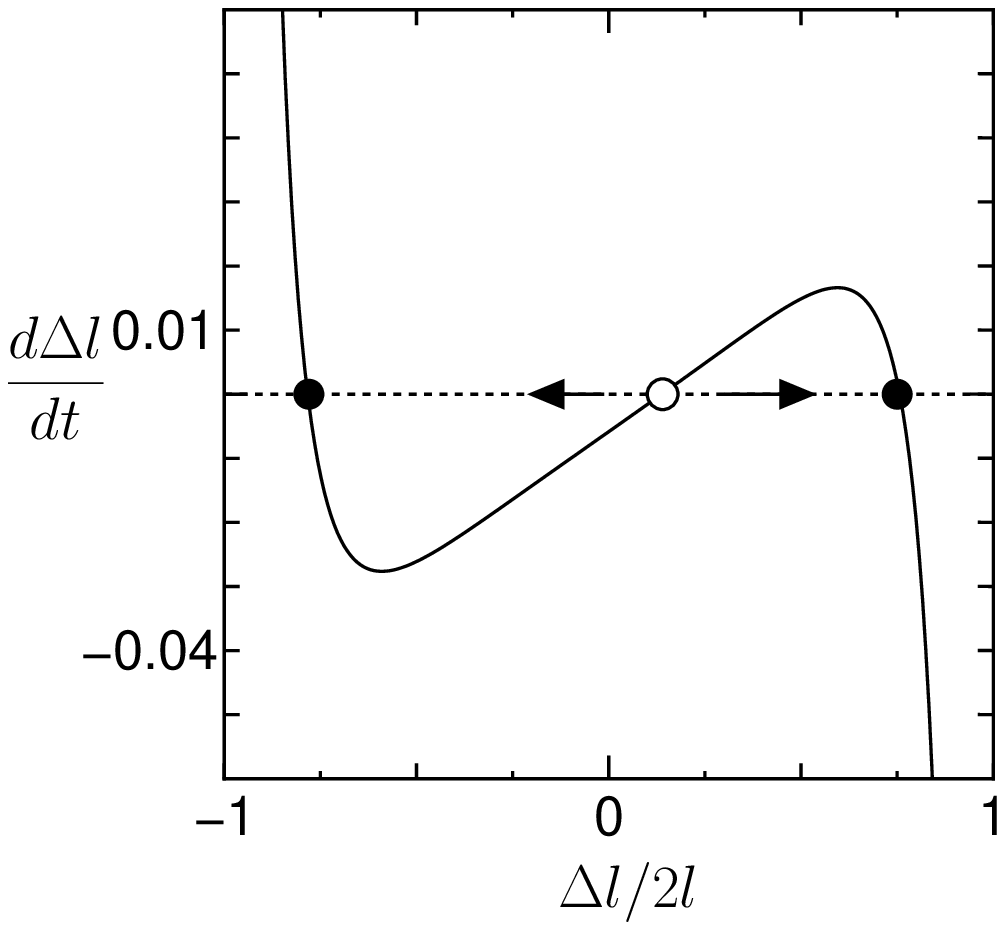}
	\centerline{\hspace*{1.2cm} {\sf \large (a)}}
	\vspace*{0.2cm}
	\\
	\includegraphics[width=0.7\linewidth]{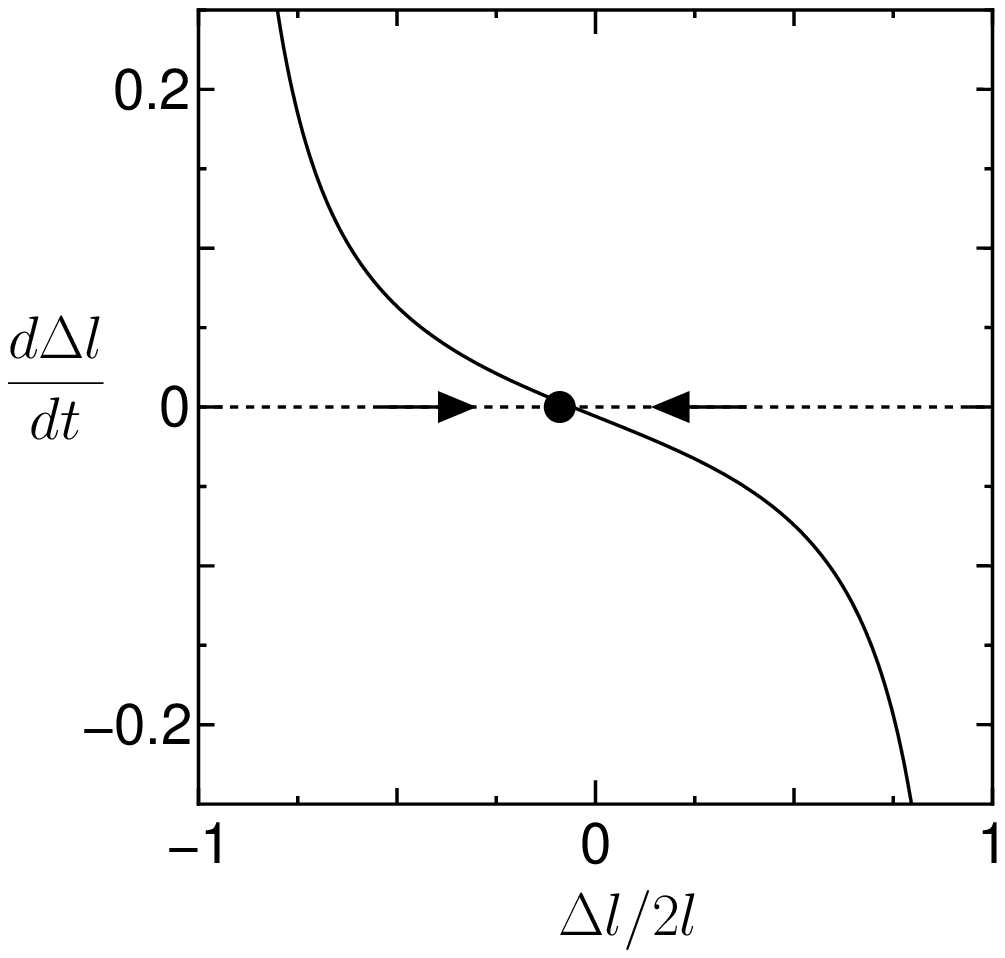}
	\centerline{\hspace*{1.2cm} {\sf \large (b)}}
	\caption{
	Time derivative of the terrace width change  with 
	(a)  step-down drift and (b) step-up drift.
	Parameters are the same as in Fig.~\ref{fig:alternating-kinetics1}
	}\label{fig:deltak}
	\end{figure}

When the kinetic coefficient $K_B$ is large,
$D_\mathrm{s}/K_\mathrm{B} \ll l$,
time evolution of step positions is shown in Fig.~\ref{fig:alternating-kinetics1}.
With step-down drift (Fig.~\ref{fig:alternating-kinetics1}(a)), 
most steps form pairs and large $T_\mathrm{A}$ terraces appear in the initial 
stage,
but formation of triplets  also occurs 
and a large $T_\mathrm{B}$ is produced.
In a late stage, 
all large terraces  become $\mathrm{T_A}$ after coalescence of small bunches.
With step-up drift (Fig.~\ref{fig:alternating-kinetics1}(b)),
the pairing does not seem to occur.

Figure~\ref{fig:deltak} represents 
the growth rate of $\Delta l$,
which is defined in the same way as in Sec.~\ref{sec:sbinc}.
With step-down drift,
an unstable fixed point with a small positive $\Delta l$
and two stable fixed points with  a large amplitude of $|\Delta l |$ 
are present.
From the condition $V_m =0$ in eq.~(\ref{eq:vmful}),
the difference of terrace width at the fixed point
near $\Delta l = 0$  is given by
	\begin{equation}
	 \frac{\Delta l}{ l}
	= \frac{K_A-K_B}{K_A+ K_B} fl,
	\label{eq:deltalk}
	\end{equation}
where we have used $fl \ll 1$ and neglected the step-step repulsion.
In eq.~(\ref{eq:dlc1})
$\Delta l$ is determined only by structural parameters
and independent of the drift,
but in eq.~(\ref{eq:deltalk})
$\Delta l$ depends on the drift
and changes the sign with the drift direction.
Expect for the sign of $\Delta l$ in the unstable fixed points,
the form of $d \Delta l/d t$ is the same as
 in Fig.~\ref{fig:short1}.
The  initial vicinal face moves to the steady state
with the large negative $\Delta l$
if the initial fluctuation is small,
but  it can move to the other steady state
with large $\Delta l$
if the initial fluctuation is large.
In Fig.~\ref{fig:alternating-kinetics1}(a),
formation of a large $T_\mathrm{B}$ in the initial stage
is due to a large fluctuation of terrace width.

With step-up drift,
there is only one fixed point  with a negative $\Delta l$.
The  steady state is stable.
Since the difference of terrace width $\Delta l$ is small,
the surface looks like the original vicinal face
in Fig.~\ref{fig:alternating-kinetics1}(b).

With the alternation of diffusion coefficient,
the form of $d \Delta l/d t$ changes with decreasing
the kinetic coefficients (Figs.~\ref{fig:short2} and \ref{fig:short1}).
On the other hand,
with the alternation of kinetic coefficient,
the form of $d \Delta l/d t$ does not change even if
$D_s /K$ changes.
Thus the bunching behavior is insensitive to the ratio $D_s/K l$.

\subsection{Step bunching induced by  the evaporation}
If the evaporation is present
and the drift is absent,
the step velocity is given by
	\begin{eqnarray}
	& &
	 \frac{x_\mathrm{s} V_{m}}{\Omega D_\mathrm{s}} 
	=
	\nonumber \\
	& &
	-
	\frac{
	 (\lambda_{m+1} \sinh l_m/x_\mathrm{s} 
	+ \cosh l_m/x_\mathrm{s})c_n -c_{n+1}
	}{h_{m+1}(l_m)}
	\nonumber \\
	 &&
	-
	\frac{
	 (\lambda_{m-1} \sinh l_{m-1}/x_\mathrm{s} 
	+ \cosh l_{m-1}/x_\mathrm{s})c_n -c_{n-1}
	}{h_{m}(l_{m-1})}
	,
	\label{eq:vm}
	\end{eqnarray}
where $ \lambda_m = D_\mathrm{s} /K_m x_\mathrm{s}$,
and $h_m(l)$ is 
	\begin{eqnarray}
	h_m(l)
	&=&
	 (\lambda_{m-1}\lambda_m +1) \sinh l/x_\mathrm{s}
	\nonumber \\
	& &
	+ (\lambda_m + \lambda_{m-1}) \cosh l/x_\mathrm{s}.
	\end{eqnarray}
Due to the evaporation, both $S_\mathrm{A}$ steps
and $S_\mathrm{B}$ steps recede.
When the steps are  equidistant,
the difference of step velocities
is given by
	\begin{eqnarray}
	 V_{2n} -V_{2n-1}
	 =
	 \frac{(\lambda_\mathrm{A} -\lambda_\mathrm{B}) \tanh  l/x_\mathrm{s}
	 }{
	 (1+ \lambda_\mathrm{A} \lambda_\mathrm{B}) \tanh l/x_\mathrm{s} 
	+ (\lambda_\mathrm{A} + \lambda_\mathrm{B})
	}.
	\end{eqnarray}
Since $ \lambda_\mathrm{A}$ is smaller than  $\lambda_\mathrm{B}$,
$S_\mathrm{B}$ recedes faster than $\mathrm{S}_\mathrm{A}$.
The vicinal face is unstable,
and large $\mathrm{T_B}$ terraces
and small $\mathrm{T_A}$ terraces appear by step pairing. 

	\begin{figure}[htp]
	\includegraphics[width=0.7\linewidth]{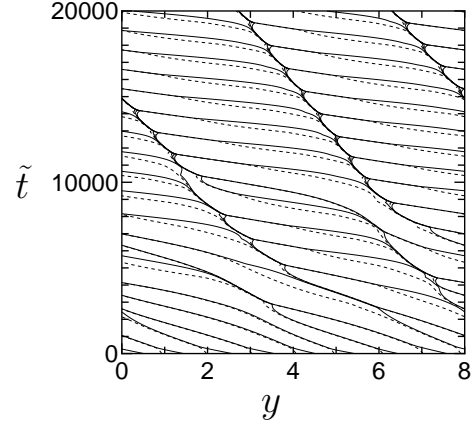}
	\caption{
	Time evolution of step position.
	$x_\mathrm{s}=15$, $K_\mathrm{A}=0.125$, $K_\mathrm{B}=1.0$,
	$D_\mathrm{s}=1.0$.
	The number of step is 16 and the system size is 8
	 with the periodic boundary condition. 
	$\tilde{A}_{\tilde{\nu}} = \times 10^{-4}$ with $\tilde{\nu} =1$.
	}\label{fig:e-keva}
	\end{figure}

Figure~\ref{fig:e-keva} represents
the time evolution of step positions,
which is obtained by numerical integration of eq.~(\ref{eq:vm}).
In the simulation,
the difference of kinetic coefficients is very large.
In very early stage,
pairing of steps occurs
and large $T_\mathrm{B}$ terraces appear.
Via coalescence of step pairs,
step bunches  appear.
The type of large terrace is the same as that in the step bunching
induced by the drift.
In large terraces, 
isolated step pairs  are present.
When a collision between a step bunch and  a step pair occurs,
another step pair leaves from the upper side of the step bunch.
A step pair  breaks into single steps,
but only temporarily.
Repeating of such collisions is  seen 
in other bucnhing systems~\cite{Sato-u95prb,Sato-u99ss}.
Here  a step pair is the fundamental unit.

\section{Summary and Discussion}\label{sec:summary}
In this paper, we studied the drift-induced step bunching with the alternation
of structural parameters:
the diffusion coefficient and the  kinetic coefficient.

With the alternation of diffusion coefficients
on consecutive terraces,
the step bunching occurs irrespective of the drift direction
if the kinetic coefficients are large.
The type of large 
terraces  is determined by the drift direction.
When the kinetic coefficients are small, 
the step bunching occurs with step-down drift
and the initial fluctuation of step distance influences 
the type of step pairs.
The growth rate of bunches is much faster with the step-down drift, 
but the difference of growth rates decreases with the evaporation.
Without drift of adatoms, the evaporation does not induce bunching.

With the alternation of kinetic coefficient at consecutive steps, 
the step bunching occurs with the step-down drift
and does not occur with the step-up drift.
The type of large terraces is influenced by the initial 
fluctuation of step distance.
The evaporation induces step bunching even if the drift is absent,
in contrast to the diffusion coefficient case.

In the experiments~\cite{livin,latyshev-livin-aseev},
the type of terraces between bunches changes with the drift direction
when the initial step distance is large.
From our result we may conclude that 
the alternation of diffusion coefficient is essential to the bunching 
and that of kinetic coefficient is not so important. 
Also the kinetic coefficients are large if the steps are impermeable. 
Impermeable steps with large kinetic coefficients are effectively equivalent to the permeable steps~\cite{Sato-us02jcg,Sato-us00prb}.
However, 
when the step distance is smaller than the critical value,
the step bunching occurs only 
with the
 step-up current in the experiment~\cite{livin},
which is not explained by our model.
In the previous studies~\cite{Sato-us00prb,Stoyanov98ss},
the dift direction to cause the step bunching on a Si(111) vicinal face
is affected by the step permeability.
In a similar way, the disagreement in a Si(001) vicinal face may be explained by 
the step permeability, 
which remains to be explored. 

The difference of growth rate of step bunches as the change of current
direction is very small in the experiment~\cite{latyshev-livin-aseev}.
The evaporation may play an important role,
but the difference does not vanish
only with the evaporation.
In the Monte Carlo simulation~\cite{Sato-us02jcg}, 
in which steps are permeable, 
a similar behavior has been observed.
We still do not undestand the origin of such a large difference.

\begin{acknowledgments}
M.S. thanks to N. Miura for helping simulations.
This work was supported by a Grant-in-Aid from  the Japanese Society
for the Promotion of Science.
\end{acknowledgments}

\end{document}